\documentstyle[emulateapj,epsf]{article}

\def\agt{\mathrel{\raise.3ex\hbox{$>$}\mkern-14mu\lower0.6ex\hbox{$\sim$}}}
\def\alt{\mathrel{\raise.3ex\hbox{$<$}\mkern-14mu\lower0.6ex\hbox{$\sim$}}}

\newcommand{\beq}{\begin{equation}}
\newcommand{\eeq}{\end{equation}}
\newcommand{\beqn}{\begin{eqnarray}}
\newcommand{\eeqn}{\end{eqnarray}}
\newcommand{\pa}{\partial}

\newcommand{\varep}{\varepsilon}

\begin{document}

\title{Collapse of Rotating Supramassive Neutron Stars to 
Black Holes: Fully General Relativistic Simulations}

\author{Masaru Shibata \altaffilmark{1}}

\affil
{\altaffilmark{1} 
Graduate School of Arts and Sciences, 
University of Tokyo,
\break
Komaba, Meguro, Tokyo 153-8902, Japan}



\begin{abstract}

We study the final state of the gravitational collapse of 
uniformly rotating supramassive neutron stars 
by axisymmetric simulations in full general relativity.
The rotating stars provided as the initial condition 
are marginally stable against quasiradial gravitational collapse 
and its equatorial radius 
rotates with the Kepler velocity (i.e., the
star is at the mass-shedding limit). 
To model the neutron stars, we adopt the 
polytropic equations of state for a wide range of 
the polytropic index as $n=2/3$, 4/5, 1, 3/2 and 2. 
We follow the formation and evolution of the
black holes, and show that irrespective of the value of
$n~(2/3\leq n \leq 2)$, the final state is a Kerr black hole and 
the disk mass is very small ($< 10^{-3}$ of the initial stellar mass). 

\end{abstract}




\section{Introduction}

Neutron stars are in general rotating.  Rotation can support 
neutron stars with higher mass than the maximum static limit, producing
supramassive stars, as defined and numerically computed by Cook et al. 
(1992, 1994a). Supramassive neutron stars may be created 
(i) when neutron stars accrete gas from a normal binary companion 
(Cook et al. 1994b), 
(ii) after the merger of binary neutron stars 
(Shibata \& Ury\=u 2000, 2002), and
(iii) after gravitational collapse of massive stellar core. 

Since viscosity drives any equilibrium star to a uniformly rotating state, 
stationary neutron stars are believed to be uniformly rotating.
The final state after the collapse of the marginally stable and  
uniformly rotating supramassive neutron stars  
is the subject of this paper. 

Rotating neutron stars with a density higher than a critical value 
are unstable against gravitational collapse. 
Such critical density is determined using the turning point 
theorem (Friedman et al. 1988; Cook et al. 1992). 
The final state of the unstable spherical stars in the adiabatic collapse 
is a Schwarzschild black hole. 
On the other hand, in the rotating case, it is not trivial:
All the fluid elements may not collapse to a Kerr black hole, leaving 
a fraction of the mass around the black hole to form disks. 
The final state after the collapse of rotating stars is
one of the fundamental questions in general relativistic astrophysics. 

To clarify the final state of the gravitational collapse of 
rotating neutron stars, numerical simulations in full 
general relativity are the best approach. 
Two groups have already performed the simulations for relativistic collapse
of rotating stars (Nakamura 1981; Nakamura et al. 1987;
Stark \& Piran 1985; Piran \& Stark 1986). However, they have not 
studied the collapse of marginally 
stable rotating neutron stars which 
are plausible initial conditions for the collapse in nature.
Probably this is because numerical 
methods for computation of initial data sets describing rapidly rotating
neutron stars, as well as numerical tools, techniques and sufficient
computational resources have become available only quite recently.
Over the last 15 years, robust numerical techniques for 
constructing equilibrium models of rotating neutron stars in full 
general relativity 
have been established (Komatsu et al. 1989; Cook et al. 1992;
Salgado et al. 1994; Stergioulas 1998). More recently, robust 
methods for the numerical evolution of the coupled equations of 
Einstein's and hydrodynamic equations have been also established 
(e.g., Shibata 1999b, 2003; Font 2000; Font et al. 2002;
Siebel et al. 2002, 2003). 

In a previous paper (Shibata et al.~2000), 
we reported the first numerical result for 
the gravitational collapse, which was 
computed by a three-dimensional numerical implementation 
in full general relativity. In that paper, we adopted 
the polytropic equation of state with $n=1$ where
$n$ is the polytropic index, and gave a 
uniformly rotating and marginally stable neutron star 
at a mass-shedding limit 
(at which the equator of a star rotates with the Kepler velocity) 
as the initial condition.
The total grid number in the simulations was only
$153 \times 77 \times 77$ for $x-y-z$ (we assumed 
the equatorial plane symmetry and $\pi$ rotation symmetry) 
because of the restricted computational resources at that time, 
and as a result, the equatorial radius (polar radius) 
of the neutron star is covered only by 40 (23) grid points initially. 
We found that the collapse leads to a black hole (we determined 
the location of the apparent horizon), and indicated 
that nonaxisymmetric instabilities do not turn on during the collapse.
However, we were not able to determine the final state
of the gravitational collapse because of the insufficient 
grid resolution. 

Since nonaxisymmetric instabilities are not likely to be relevant during 
the collapse, the simulation should be carried out under the assumption of 
the axial symmetry. With this restriction, we could significantly improve 
the grid resolution for a given computational resource. 
Motivated by this fact, we have constructed a numerical code for 
axisymmetric numerical simulation in full general relativity,
which has been already completed (Shibata 2000, 2003). 
Because of the restriction to the axial symmetry as well as
progress in computational resources, we can 
easily increase the grid number 3--5 times as large as 
that in the previous three-dimensional simulation (Shibata et al. 2000)
even in inexpensive personal computers. 
As a result, we can search for convergent 
numerical results changing the grid number for a wide range
with inexpensive computational cost. 
In addition, we adopt a high-resolution shock-capturing scheme 
for evolving the hydrodynamic equations (Shibata 2003), which enables 
us to assess whether shocks play an important role during 
the collapse to a black hole.

In this paper, we present new numerical results for gravitational collapse 
computed by the new axisymmetric numerical implementation. The simulations
were carried out setting marginally stable equilibrium
neutron stars as the initial condition. We focus on the collapse of 
uniformly rotating supramassive neutron stars at mass-shedding limits
as before. 
By exploring rotating stars at mass-shedding limits, we can 
clarify the final state of the collapsed objects most efficiently. 
To investigate the effect of the stiffness of equations of state, 
we adopt polytropic 
equations of state with a wide variety of the polytropic index. 
The state of marginally stable stars which is characterized by the 
compactness, angular momentum parameter and 
density distribution depends strongly on the equations of state. 
This implies that the final state after the gravitational collapse
of rotating stars could depend strongly on the
equations of state, in contrast to the collapse of nonrotating stars
in which the Schwarzschild black hole with no disks is the unique outcome. 
To classify the type of gravitational collapse and its final state, 
a systematic study for a wide variety of
equations of state is essential. 

In Sec. 2, we briefly describe our formulation, initial data, and
spatial gauge conditions.  In Sec. 3, we present numerical results.
In Sec. 4, we provide a summary.  Throughout 
this paper, we adopt the units $G=c=K=1$ where $G$, $c$ and
$K$ denote the gravitational constant, speed of light and
polytropic constant, respectively.  We use Cartesian coordinates $x^k=(x, y,
z)$ as the spatial coordinate with $r=\sqrt{x^2+y^2+z^2}$;
$t$ denotes coordinate time. 

\section{Summary of formulation and initial conditions} 

We performed hydrodynamic simulations in general relativity 
for the axisymmetric spacetime using the same formulation as 
that used in a previous paper (Shibata 2003), 
to which the reader may refer for details and basic equations.

We assume that neutron stars are composed of
the inviscid, ideal fluid. Then, the 
fundamental variables for the hydrodynamic equations are: 
\beqn
\rho &&:{\rm rest~ mass~ density},\nonumber \\
\varep &&: {\rm specific~ internal~ energy}, \nonumber \\
P &&:{\rm pressure}, \nonumber \\
u^{\mu} &&: {\rm four~ velocity}, \nonumber \\
v^i &&={dx^i \over dt}={u^i \over u^t},
\eeqn
where subscripts $i, j, k, \cdots$ denote $x, y$ and $z$, and
$\mu$ the spacetime components. 
As the fundamental variables to be evolved
in the numerical simulations, 
we in addition define a density $\rho_*(=\rho \alpha u^t e^{6\phi})$
($\phi$ is defined below)
and weighted four-velocity $\hat u_i (= (1+\varepsilon+P/\rho)u_i)$ from
which the total rest mass and angular momentum
of the system can be integrated as
\beqn
M_*&=&\int d^3 x \rho_*, \\
J  &=&\int d^3 x \rho_*\hat u_{\varphi}. 
\eeqn
General relativistic hydrodynamic equations are solved using 
the so-called high-resolution shock-capturing scheme
(Shibata 2003; see Font 2002 for a general review
of high-resolution shock-capturing schemes)
with the cylindrical coordinates. 

We neglect effects of viscosity and magnetic fields.
The dissipation and angular momentum transport timescales
by these effects are much longer than the dynamical timescale
unless the magnitude of viscosity and magnetic fields is
extremely large (e.g., Baumgarte et al. 2000).
Thus, neglecting them is appropriate assumption.

The fundamental variables for the geometry are: 
\beqn
\alpha &&: {\rm lapse~ function}, \nonumber \\
\beta^k &&: {\rm shift~ vector}, \nonumber \\
\gamma_{ij} &&:{\rm metric~ in~ 3D~ spatial~ hypersurface},\nonumber \\ 
\gamma &&=e^{12\phi}={\rm det}(\gamma_{ij}), \nonumber \\
\tilde \gamma_{ij}&&=e^{-4\phi}\gamma_{ij}, \nonumber \\
K_{ij} &&:{\rm extrinsic~curvature}. 
\eeqn
As in the series of our papers, 
we evolve $\tilde \gamma_{ij}$, $\phi$, 
$\tilde A_{ij} \equiv e^{-4\phi}(K_{ij}-\gamma_{ij} K_k^{~k})$
together with the three auxiliary functions
$F_i\equiv \delta^{jk}\pa_{j} \tilde \gamma_{ik}$ and the 
the trace of the extrinsic curvature $K_k^{~k}$ 
with a free evolution code 
(see Shibata \& Ury\=u 2002 for the latest version of the
formulation).

The Einstein equations are solved in the Cartesian coordinates. 
To impose the axisymmetric boundary condition, 
the so-called Cartoon method is used (Alcubierre et al. 2001b): 
Assuming a reflection symmetry with 
respect to the $z=0$ plane, we perform simulations 
using a fixed uniform grid with the 
size $N \times 3 \times N$ in $x-y-z$ which covers
a computational domain as
$0 \leq x \leq L$, $0\leq z \leq L$ and $-\Delta \leq y \leq \Delta$. 
Here, $N$ and $L$ are constants and $\Delta = L/N$. 
For $y=\pm \Delta$, the axisymmetric boundary conditions are imposed.

The slicing conditions are 
basically the same as those adopted in previous papers
(Shibata 1999, 2000, 2003; Shibata \& Ury\=u 2000, 2002); i.e., we
impose an approximate maximal slice condition ($K_k^{~k} \simeq 0$).
On the other hand, 
we adopt two spatial gauge conditions for the shift vector. 
One is an approximate minimal distortion (AMD)
gauge condition [$\tilde D_i
(\pa_t \tilde \gamma^{ij}) \simeq 0$ where $\tilde D_i$ is the
covariant derivative with respect to $\tilde \gamma_{ij}$]
(Shibata 1999) which has been used in our previous works.
In contrast with previous papers (e.g., Shibata et al. 2000), 
we used the AMD gauge condition without modification. 
The other is a dynamical gauge condition
(e.g., Alcubierre et al. 2001a; Lindblom \& Scheel 2003).
In the present work, we impose the dynamical gauge condition with 
the equation
\beq
\pa_t \beta^k = \tilde \gamma^{kl} (F_l +\Delta t \pa_t F_l),
\label{dyn}
\eeq
where $\Delta t$ denotes a timestep in numerical computation.
The second term in the right-hand side of Eq. (\ref{dyn})
is introduced to stabilize numerical computation. 
With this choice, $\beta^k$ obeys a hyperbolic-type equation
(for a sufficiently small value of $\Delta t$), because 
the right-hand side of the evolution equation for $F_l$ 
contains vector Laplacian terms as $\beta^k_{~,ii}+\beta^{i}_{~,ik}/3$ 
(e.g. Shibata \& Nakamura 1995; Shibata \& Ury\=u 2002). 
The outstanding merit of this gauge condition 
is that we can save computational time
significantly, since we do not have
to solve elliptic-type equations. 
In the numerical computations, we adopted these two spatial 
gauge conditions, and 
found that both give the (almost) identical numerical results: 
As in the case of the AMD gauge condition, 
the dynamical gauge enables to carry out a longterm stable simulation
irrespective of the equations of state. 
Thus, here, we present numerical results 
with the dynamical gauge condition to demonstrate its robustness. 

During numerical simulations, 
violations of the Hamiltonian constraint and conservation of
mass and angular momentum are monitored as code checks.  Several test
calculations, including stability and collapse 
of spherical and rotating neutron stars, as well as convergence tests, 
have been described in a previous paper (Shibata 2003).
Formation of a black hole is determined by finding an apparent horizon. 

To model supramassive neutron stars, we adopted the polytropic equations of
state of the form 
\beq
P=K \rho^{1+{1\over n}}. 
\eeq
In this paper, we choose $n=2/3$, 4/5, 1, 3/2 and 2 to systematically study 
the effects of stiffness of equations of state.
During the simulations, we use a $\Gamma$-law equation of state as 
\beq
P=(\Gamma-1)\rho \varep,
\eeq 
where $\Gamma$ is the adiabatic constant which is 
set as $1+1/n$. In the absence of shocks, 
no heat is generated and the collapse is adiabatic, 
preserving the polytropic form of the equations of state. 
This implies that the quantity $P/\rho^{\Gamma}$ measures the
efficiency of the shock heating. 

As initial conditions, we gave marginally stable and uniformly rotating 
supramassive neutron stars at mass-shedding limits in equilibrium states. 
To induce gravitational collapse, we initially reduced the
pressure (i.e., $K$) uniformly by 0.5\% in all the simulations.
Whenever we reduce the pressure, we solve the equations for
the Hamiltonian and momentum constraints to enforce them at $t=0$. 

Marginally stable supramassive neutron stars of polytropic
equations of state with $2/3 \leq n \leq 2$
have the compactness $0.06 \leq M/R \leq 0.25$ 
(see Table 1). Typical compactness of neutron stars is considered to be 
$\sim 0.15$--0.2 (Shapiro \& Teukolsky 1983; Glendenning 1996).
Thus, the present choice of $n$ yields plausible models for 
marginally stable supramassive neutron stars.

Physical units enter the problem only through the polytropic constant
$K$, which can be chosen arbitrarily or else completely scaled out of
the problem. Thus, we only display the dimensionless quantities
which are defined as
\beqn \label{rescale}
\bar M_* = M_* K^{-n/2}, & \bar M =  M K^{-n/2}, & 
\bar R = R K^{-n/2}, \nonumber  \\
\bar J = J K^{-n},  &  \bar{\rho} = \rho K^n, & 
\bar \Omega = \Omega K^n, \label{scale}
\eeqn
where $M$, $R$, and $\Omega$ denote the ADM mass, equatorial circumferential
radius and the angular velocity.
Hereafter, we adopt the units of $K=1$ so that we will omit the bar. 

In Table 1, we list the rotating stars at 
mass-shedding limits that we picked up as initial conditions
in the present simulations. 
All the quantities are scaled to be nondimensional using
the relation described in Eq. (\ref{scale}).
Stability of uniformly rotating polytropes with $n=1$, 3/2 and 2 against
gravitational collapse has been already studied by Cook et al.(1994). 
Thus, for these polytropic indices, we choose the stars close 
to the marginally stable point 
based on their results. For $n=2/3$ and 4/5, we do not know the
critical point for the stability. 
As shown by Cook et al. (1994), however,
for stiff equations of state with 
$n \alt 1$, the stability of the uniformly rotating stars at
mass-shedding limits changes near a point where the mass is maximum.
Thus, we choose the stars of nearly maximum mass 
along the sequence of the uniformly rotating star at mass-shedding limits.

The ratio of the kinetic energy to the gravitational binding
energy for all the stars that we picked up here 
is much smaller than $0.27$ which is a widely believed critical value 
for onset of the dynamical bar-mode instability
in a uniformly rotating star (Chandrasekhar 1969).
Thus, the nonaxisymmetric deformation is unlikely to turn on
during collapse.
This justifies that we assume the axial symmetry.

\section{Numerical results}

\subsection{Prediction}

Before presenting numerical results, 
we here predict the plausible outcome of the gravitational collapse. 
Such prediction helps to understand the
reason that a result obtained in a numerical
simulation should be the output. 

Because of the axial symmetry of the system
(and since the fluid is assumed to be inviscid), 
the mass distribution as a function of the specific angular momentum
as well as the baryon rest-mass and angular momentum are conserved 
throughout the evolution of the system. Using this fact, 
we can predict the final state of gravitational
collapse from the initial condition.

We define the mass distribution as a function of the
specific angular momentum according to (e.g., Stark \& Piran 1987)
\beq
M_*(j_0)=\int_{j \leq j_0} d^3x \rho_*,
\eeq
where $j$ is a value of the specific angular momentum computed as 
$x\hat u_y(=h u_{\varphi})$ and $j_0$ denotes a particular value for $j$. 
In Fig. 1, we show the mass distribution as a
function of the specific angular momentum 
$M_*(j)$ as a function of $j/M$.

To predict the final state of the collapse, we 
assume that (i) a black hole is formed after the collapse,
(ii) most of the mass elements fall into the black hole, 
and (iii) the value of the nondimensional angular momentum
parameter $q \equiv J/M^2$ of 
a formed black hole is nearly equal to the value of the system. 

Since the value of $q$ of all the stars that we 
picked up here is smaller than unity and 
no heat source exist in the collapsing star,
assumption (i) is quite reasonable.
(ii) and (iii) are also reasonable because
the progenitor of the collapse is uniformly rotating, so that
the effect of centrifugal force which prevents a fluid element
falling into a black hole is important only around the 
low-density outer region of the collapsing stars. 

According to (i)--(iii), we assume that the mass and Kerr
parameter of the formed black hole are $\approx M$ and $\approx M q$.
Around the Kerr black hole, there is the innermost stable
circular orbit (ISCO). All the mass elements of circular orbits
inside the ISCO have to fall into
the black hole. This implies that a mass element
of the specific angular momentum which is smaller than
the value at the ISCO, $j_{\rm ISCO}$, has to fall into the black hole.

Assuming that the mass and Kerr parameter of the formed
Kerr black holes are $M$ and $M q$, we computed $j_{\rm ISCO}$
for the models listed in Table 1 
using the formula derived by Bardeen et al. (1972). 
The numerical results are described at the last column of Table 1. 
In all the models, $j_{\rm ISCO}$ is larger than $2.5M$.
(Note that it is $2\sqrt{3}M$ for $q=0$.)
From Fig. 1, we find that the fraction of the mass
with $j_{\rm ISCO} > 2.5M$ is approximately zero (less than $10^{-3}$)
irrespective of $n$. 
Therefore, the final state is predicted to be a Kerr black hole and the 
disk mass is very small ($<10^{-3}$ of the initial stellar mass) for
any value of $n$ between 2/3 and 2.

\subsection{Formation and evolution of black holes}

We performed simulations 
varying $N$ for a wide range as 180--480. This
grid number is several times larger than that in
a previous study (Shibata et al. 2000), and enables 
to check the convergence of the numerical solutions in detail. 
For $n \leq 3/2$, the equatorial radius of marginally 
stable rotating stars are initially covered by $N/2$ grid points.
The polar radius is covered by $\approx 0.3N$ in this case.
For $n=2$, equilibrium stars have a more centrally-concentrated
density configuration than that for stiffer equations of state.
To resolve the central region, we arrange the
grid spacing with which the equatorial radius is covered by $5N/6$
grid points initially.
In this case, the polar radius is covered by $\approx N/2$ grid points.

Numerical computation was in part performed on FACOM VPP5000
in the data processing center of National Astronomical 
Observatory of Japan, but most of the simulations were carried out 
using personal computers of Pentium-4 processors, 
each of which has 2 Gbytes memory and 2.8 GHz clock.  Even for $N=480$, 
it takes only about 5 days to finish a job
of $\sim 40000$ time steps on one of these computers. 

As a result of the simulations, we found that 
irrespective of the value of $n$, 
the collapse proceeds monotonically to be 
a Kerr black hole. During the collapse, shock heating 
is negligible in the central region. Namely, $P/\rho^{\Gamma}$ remains 
approximately constant. 

In all the simulations, the apparent horizons were determined 
in the late phase of the collapse. As reported in
a previous paper (Shibata 2003), 
accuracy of numerical results, measured by the
violation of the Hamiltonian constraint, deteriorates
monotonically with time, and the 
computations eventually crashed due to the grid stretching
around the black holes. 
However, the grid number inside the surface of the apparent horizons
in this work is large enough to 
resolve the formation, and evolution of the black hole 
for a duration $\sim 20M$ even without black hole excision techniques
(W. Unruh 1984, unpublished; Seidel \& Suen 1992). 
The duration is in general longer with better grid resolutions, and
with the largest grid number, we could determine the final
state of the collapse approximately. 
However, to carry out a simulation for more than $20M$ after the 
formation of black holes, excision techniques are absolutely necessary. 

We have also checked that the 
mass distribution as a function of the
specific angular momentum is conserved accurately.
In Fig. 1(b), we compare the mass distribution at $t=0$ and 
at the formation of apparent horizon for $n=2$ as an example. 
The figure shows a good conservation of it. 

In Fig. 2, we display the square of the 
mass of the apparent horizons $M_{\rm AH}$ as 
a function of time. Here, $M_{\rm AH}$ is defined as 
\beq
M_{\rm AH} \equiv \sqrt{{A \over 16\pi}}, 
\eeq
where $A$ denotes the area of the apparent horizon 
(e.g., Cook \& York 1990).
Figure 2 shows that $M_{\rm AH}$ approaches an asymptotic value.
It is also found that the numerical results are convergent 
with increase of $N$.

Together with the evolution of $M_{\rm AH}^2$, in Fig. 2, 
we plot the square of the irreducible mass of 
the event horizon $M_{\rm irr}^2$ 
for the Kerr black hole of $M$ and $J$ (dotted horizontal lines) as
\beq
M_{\rm irr}^2={1 \over 2}\Big(M^2+\sqrt{M^4-J^2}\Big). 
\eeq 
Here, as $M$ and $J$, we adopt the total values of the system 
which are computed from the initial data sets. 
If the final state of the gravitational collapse 
is a Kerr black hole with negligible disk mass, the 
mass of the apparent horizon should approach $M_{\rm irr}$. 
Figure 2 clearly shows that $M_{\rm AH}$
asymptotically approaches $M_{\rm irr}$. 
Small deviation of the asymptotic value
of $M_{\rm AH}$ from $M_{\rm irr}$
is likely to be a numerical error. Indeed,
with improvement of the grid resolution, the final value of 
$M_{\rm AH}$ appears to converge to $M_{\rm irr}$. 
With the best resolution, the magnitude of the numerical error
in $M_{\rm AH}$ is less than 1\%. 
This implies that the final state of the collapse is
the Kerr black hole and the fraction of the disk mass is
very small (within the magnitude of numerical error $\alt 1$\%). 

To reconfirm this fact, we display the evolution of the fraction of 
the mass located outside a coordinate radius $r_0$, 
$M_*(r>r_0)/M_*$, for $n=2$ (solid curve), 3/2 (dotted curve), 
1 (dashed curve), and 2/3 (long-dashed curve) in Fig. 3.
The result for $n=4/5$ is essentially the same, so that we omit it.  
Here, $r_0$ is chosen as $\approx 0.6M$, which is approximately equal 
to the asymptotic value of the 
coordinate polar-radius of the apparent horizon in the present
computations. 
We found that in our gauge, the coordinate equatorial-radius is
slightly larger than the polar-radius, so that 
the fraction of the mass outside the apparent horizon is slightly
smaller than $M_*(r>r_0)/M_*$. Figure 3 shows that 
before the collapse, most of the fluid elements are located
outside $r=r_0$, but during the collapse, 
$M_*(r>r_0)/M_*$ monotonically decreases and approaches zero. 
This confirms the results in Fig. 2. 
The present results also reconfirm the same conclusion 
reached previously in less-resolved simulations
for $n=1$ (Shibata et al. 2000). 

We note that for $n=2$, 
a small amount of mass elements appears to be outside
the black hole at the termination of the simulations. 
We infer that they will be eventually swallowed into the 
black hole because the specific angular momentum for the 
fluid element is not large enough to form disks around the black hole. 

Recall that we study the collapse of 
uniformly rotating stars of maximum angular velocity, 
implying that the effect of rotation is taken into account most efficiently. 
Therefore, we conclude that the final state after the collapse 
of {\it all} the marginally stable and uniformly rotating polytropic stars 
with $2/3 \leq n \leq 2$ is a Kerr black hole and the disk mass
is $< 10^{-3}M_*$. 

It should be noted that for smaller value of $n$, $M_{\rm AH}$ reaches
$M_{\rm irr}$ more quickly. For $n=2/3$, the growth timescale of $M_{\rm AH}$ 
from 0 to $\sim M_{\rm irr}$ is $\sim 6M$ while 
for $n=2$, it is larger than $30M$. 
This reflects that the nature of the collapse depends strongly 
on the initial density configuration which is determined by
the stiffness of the equations of state.
For stiffer equations of state,
the density of the initial condition distributes rather uniformly. 
Thus, the collapse proceeds coherently. 
For softer equations of state, on the other hand, 
the initial condition has more centrally-concentrated 
density distribution with low-density outer envelops. 
Thus, the central region collapses to a black hole earlier,
and then the outer region falls into the black hole
spending a longer timescale than that for stiffer equations of state. 

To illustrate that the identical numerical results were
obtained in two different spatial gauge conditions, 
in Fig. 4, we display evolution of
$\phi$ and $\rho_*$ at $r=0$ as well as $M_{\rm AH}^2$
for $n=3/2$ with $N=360$.
The solid and dotted curves denote the numerical results
in the dynamical and AMD gauge conditions, respectively.
We see that both results are in good agreement. 

Finally, we address the following point:
The collapse of compact stars to black holes is among the most
interesting processes leading to the production of gravitational waves.
As pointed out by Stark and Piran (1985), 
quasinormal modes of a black hole would be excited after
the formation, and as a result, gravitational waves associated with such
quasinormal-mode oscillations may be emitted. 
It is an interesting subject to clarify how large the amplitude of
gravitational waves is. From this motivation,
we tried to extract gravitational waves in the simulation,
but we were not able to do because the amplitude is likely much smaller
than the typical size of numerical noise of our present simulation.
The reason that the amplitude is very small is the following:
(i) the collapse coherently proceeds, i.e., almost all the
fluid elements collapse to form a black hole simultaneously. 
In such case, the excitation of the quasinormal modes of 
a black hole is likely to be weak, because the
quasinormal modes are excited by perturbations struck after formation
of a black hole; (ii) the nondimensional angular momentum parameter $q$ 
is not very large $< 0.7$. As Stark and Piran showed that
a large amount of gravitational waves is emitted for $q$ close to 1.

\section{summary and discussion}

We have reported new numerical results of 
axisymmetric simulations for the gravitational collapse of 
rapidly and uniformly rotating supramassive neutron stars to black holes
in full general relativity. The initial conditions 
for the neutron stars are given using 
polytropic equations of state for a wide range of 
the polytropic index as $n=2/3$, 4/5, 1, 3/2 and 2. 
The initial state of the rotating stars 
is marginally stable against the quasiradial gravitational collapse 
and at the mass-shedding limit. The hydrodynamic 
simulations were carried out using a high-resolution shock-capturing
scheme with the $\Gamma$-law equations of state. We have
demonstrated that irrespective of the value of $n~(2/3\leq n \leq 2)$,  
the collapse monotonically proceeds with negligible shock heating, and 
the final state is a Kerr black hole with a
small fraction of the disk mass. 

As mentioned in Sec. 3.1, 
the results obtained in this paper can be expected from
the initial conditions. In the same manner, we can 
expect the final states of the gravitational collapse 
for softer equations of state with $n > 2$.
With a large value of $n \sim 3$, we may model an unstable 
massive stellar core at the final stage of stellar evolution
and a supermassive star of $M \agt 10^5M_{\odot}$. In Fig. 5, 
we show the mass distribution as a function of the specific angular momentum 
of the marginally stable and uniformly rotating 
stars at mass-shedding limits for $n=2.5$, 2.9, and 3. 
The marginally stable stars for these polytropic indices have been already 
determined by Cook et al. (1994) for $n=2.5$ and 2.9 and 
Baumgarte and Shapiro (1999) for $n=3$. 
The nondimensional angular momentum parameter $q$ is 
$\approx 0.39$, 0.57, and 0.96 for $n=2.5$, 2.9 and 3, so that 
$j_{\rm ISCO}/M$ for a black hole of mass $M$ and angular momentum $J$ 
is $\approx 3.0$, 2.8, and 1.8, respectively. 
From Fig. 5, we can expect that the final state after the collapse 
for $n=2.5$ is a Kerr black hole and only a small fraction 
of the initial stellar elements ($\sim 10^{-3}M_*$) forms the disks. 
On the other hand, disks of mass of $\agt 0.01M_*$ and $\agt 0.1M_*$ 
are likely to be formed for $n=2.9$ and 3, respectively. 
(If the mass of the black holes is smaller than $M$, $j_{\rm ISCO}$
is also smaller and, hence, the disk mass could be larger. 
This implies that the numerical fraction mentioned here is 
the minimum value.) 
The same conclusion for $n=3$ has been already drawn by 
Shibata and Shapiro (2002) and 
Shapiro and Shibata (2002) in a more careful analysis. 

The reason why disks are formed 
for $n \agt 2.9$ is simply that the marginally stable stars 
with polytropic equations of state of such large value of $n$ have a large 
equatorial radius with $R/M \agt 200$, and hence 
the specific angular momentum for a certain fraction of 
the fluid elements is large enough to escape from swallowing into a 
black hole. 
The present study together with the previous one (Shibata \& Shapiro 2002) 
shows that nature of the collapse of rapidly rotating stars to a black hole 
depends strongly on the equations of state in particular for $n \sim 3$. 

\acknowledgments

This work was supported by Japanese 
Monbukagakusho Grants (Nos. 13740143, 14047207, 15037204 and 15740142).


\begin{table}[t]
\begin{center}
\caption{Parameters of the initial conditions}
\begin{tabular}{cccccccccc}
\tableline\tableline
$n$ & $\rho_c$ & $M_*$ & $M$ & $M/R$ & $\Omega$ & $J/M^2$
& $T/|W|$ & $\alpha(r=0)$ & $j_{\rm ISCO}/M$
\\ \tableline
2/3 & 0.730 & 0.184 & 0.158 & 0.248 & 0.791 & 0.670 & 0.117 & 0.329 & 2.64
\\ \tableline
4/5 & 0.520 & 0.190 & 0.168 & 0.212 & 0.572 & 0.626 & 0.102 &0.382 & 2.71
\\ \tableline
1   & 0.296 & 0.206 & 0.188 & 0.175 & 0.392 & 0.561 & 0.0809 &0.442 & 2.82
\\ \tableline
3/2 & 0.0570 & 0.304 & 0.290 & 0.106 & 0.117 & 0.450 & 0.0465 & 0.607 & 2.97
\\ \tableline
2 & 0.00523 & 0.559 & 0.549 & 0.0643 & 0.0242 & 0.388 & 0.0268 & 0.750 & 3.05
\\ \tableline
\end{tabular}
\tablecomments{The central density $\rho_c$,
baryon rest-mass $M_*$, ADM mass $M$, compactness $M/R$,
angular velocity, angular momentum $J$ in units of $M^2$,
ratio of the kinetic energy to the gravitational
binding energy, and central value of the lapse function. 
Here, $R$ denotes the circumference radius at the equatorial
surface. The last column shows the 
specific angular momentum of a test particle orbiting
a Kerr black hole of mass $M$ and angular momentum $J$. 
All the quantities are shown in units of $c=G=K=1$.
}
\end{center}
\end{table}

\begin{figure}[htb]
\begin{center}
\epsfxsize=3.0in
\leavevmode
(a)\epsffile{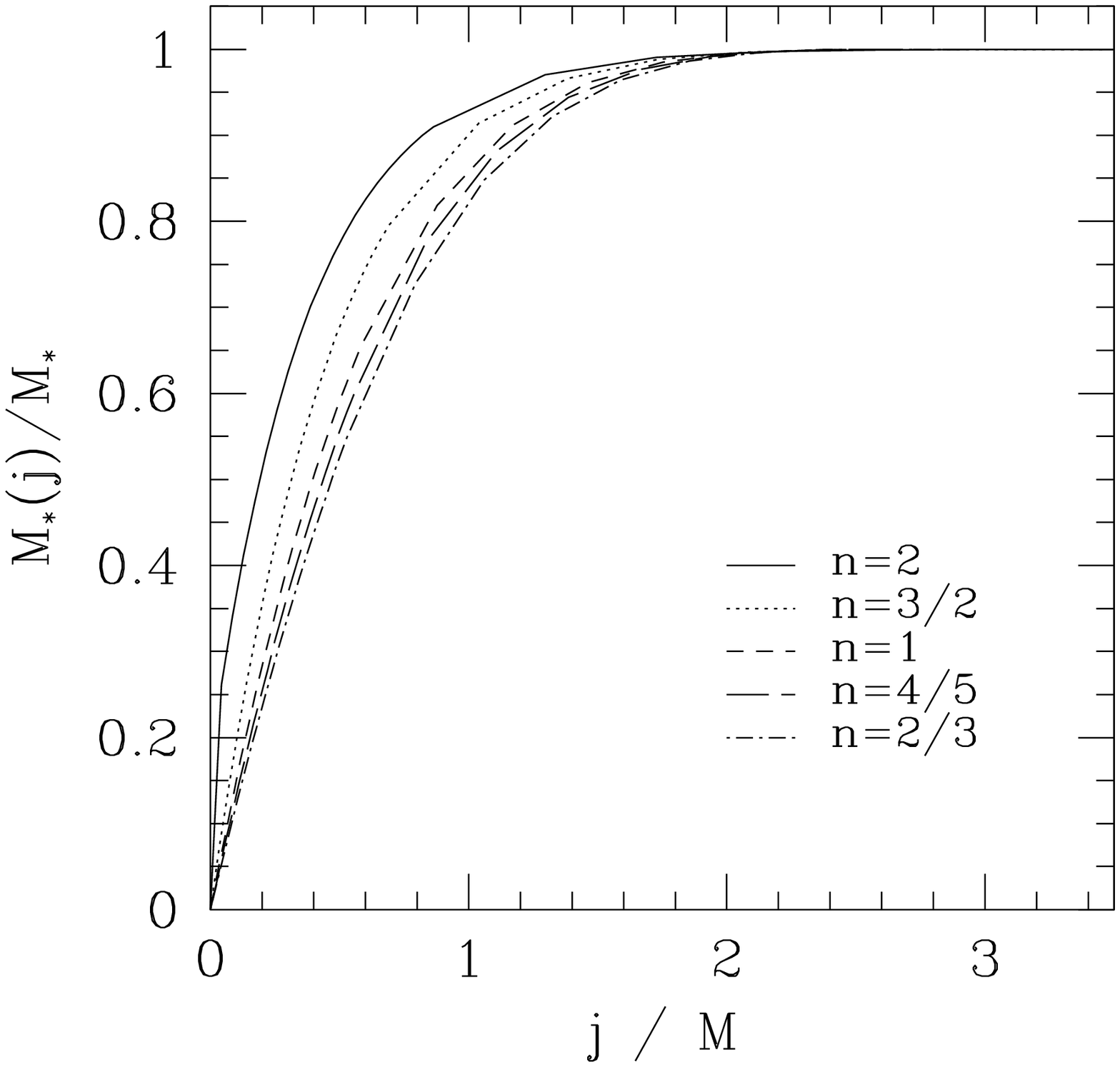}
\epsfxsize=3.0in
\leavevmode
~~~(b)\epsffile{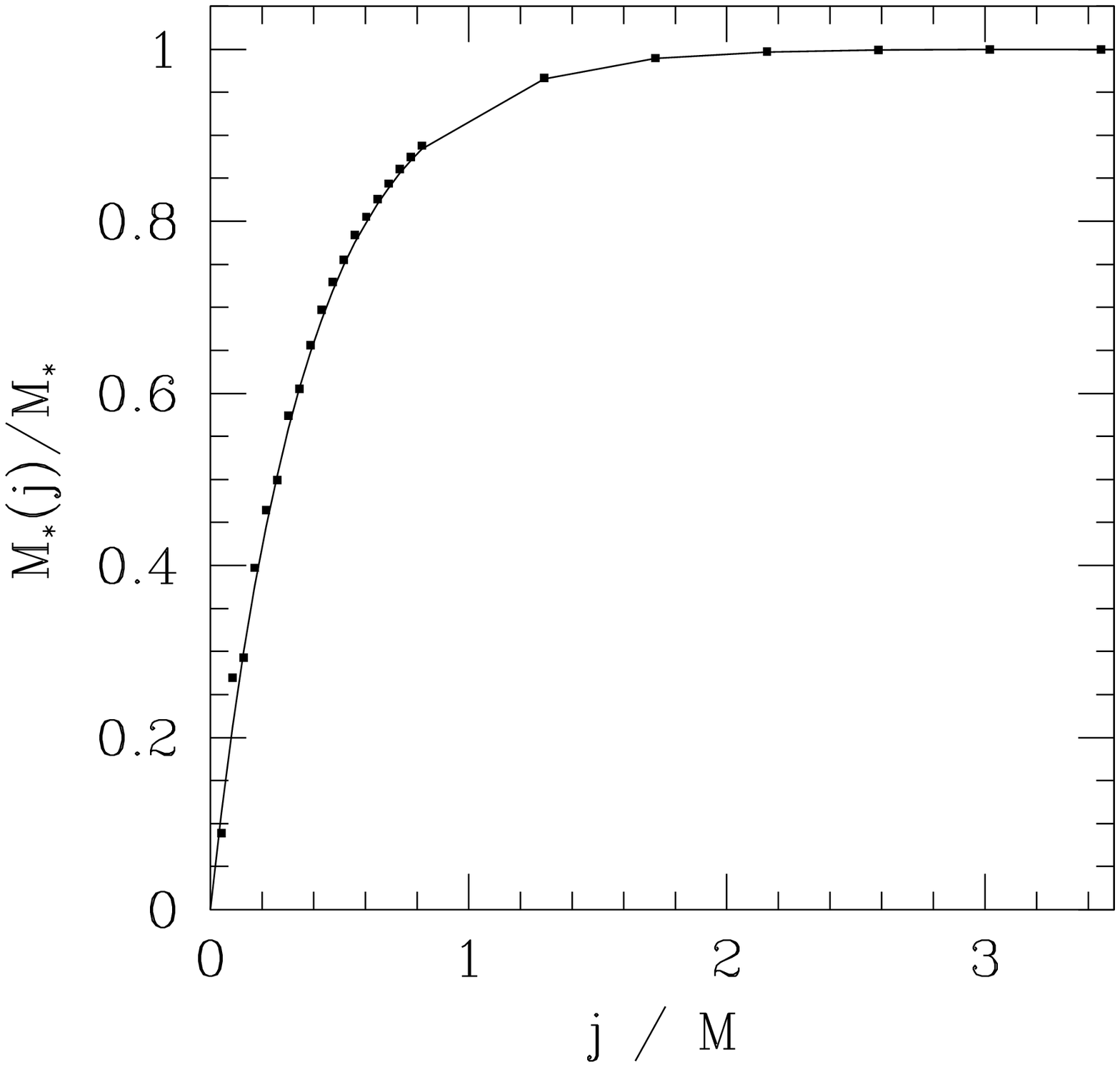}
\caption{(a) The mass distribution as a function of the
specific angular momentum for $n=2/3$--2 at $t=0$.
(b) The same as (a) but at $t=0$ (solid curves) and
at the formation of the apparent horizon (filled circles)
for $n=2$. This is the result for a simulation with $N=480$. 
\label{FIG1}
}
\end{center}
\end{figure}

\begin{figure}[htb]
\begin{center}
\epsfxsize=2.6in
\leavevmode
(a)\epsffile{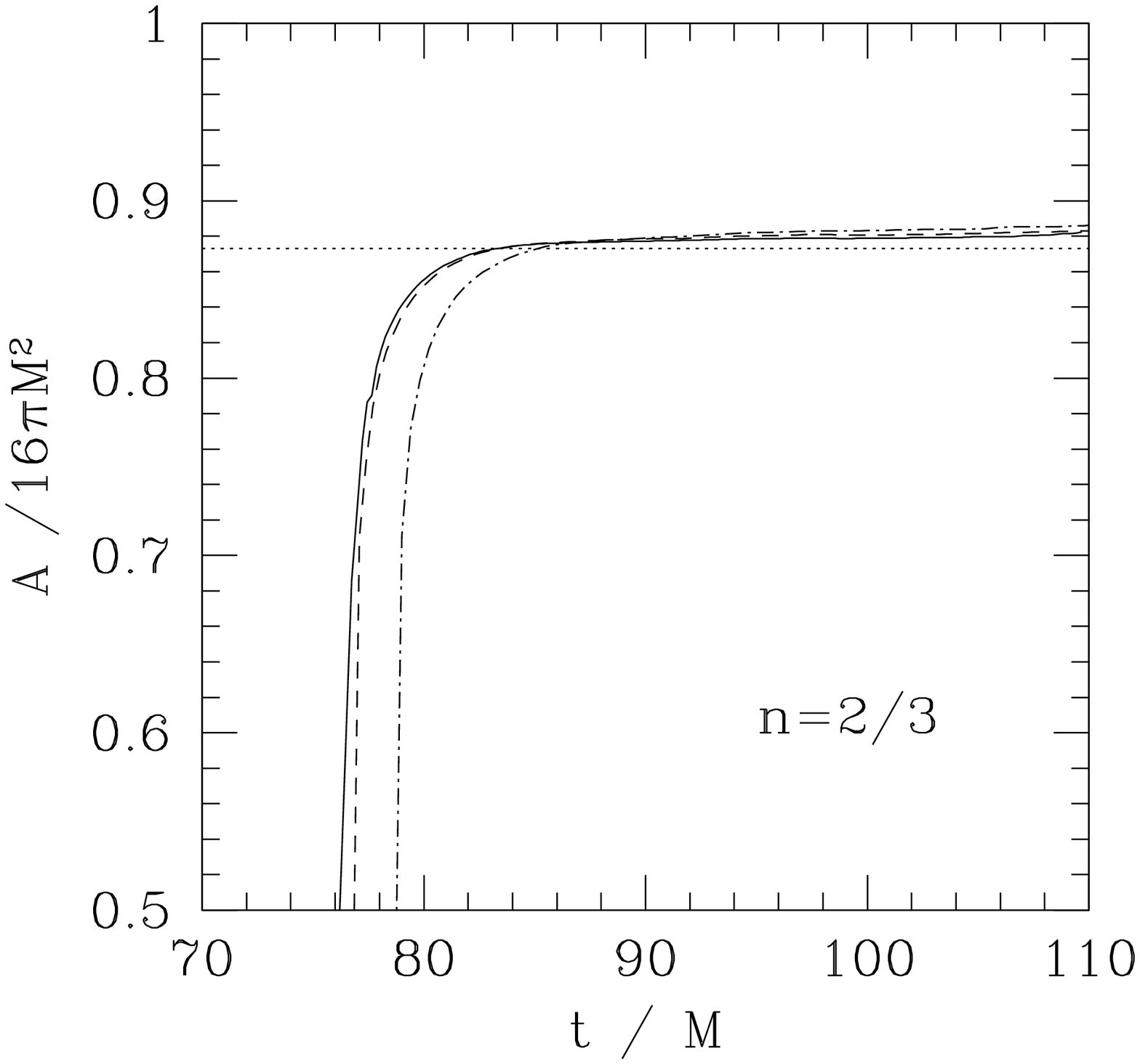}
\epsfxsize=2.6in
\leavevmode
~~(b)\epsffile{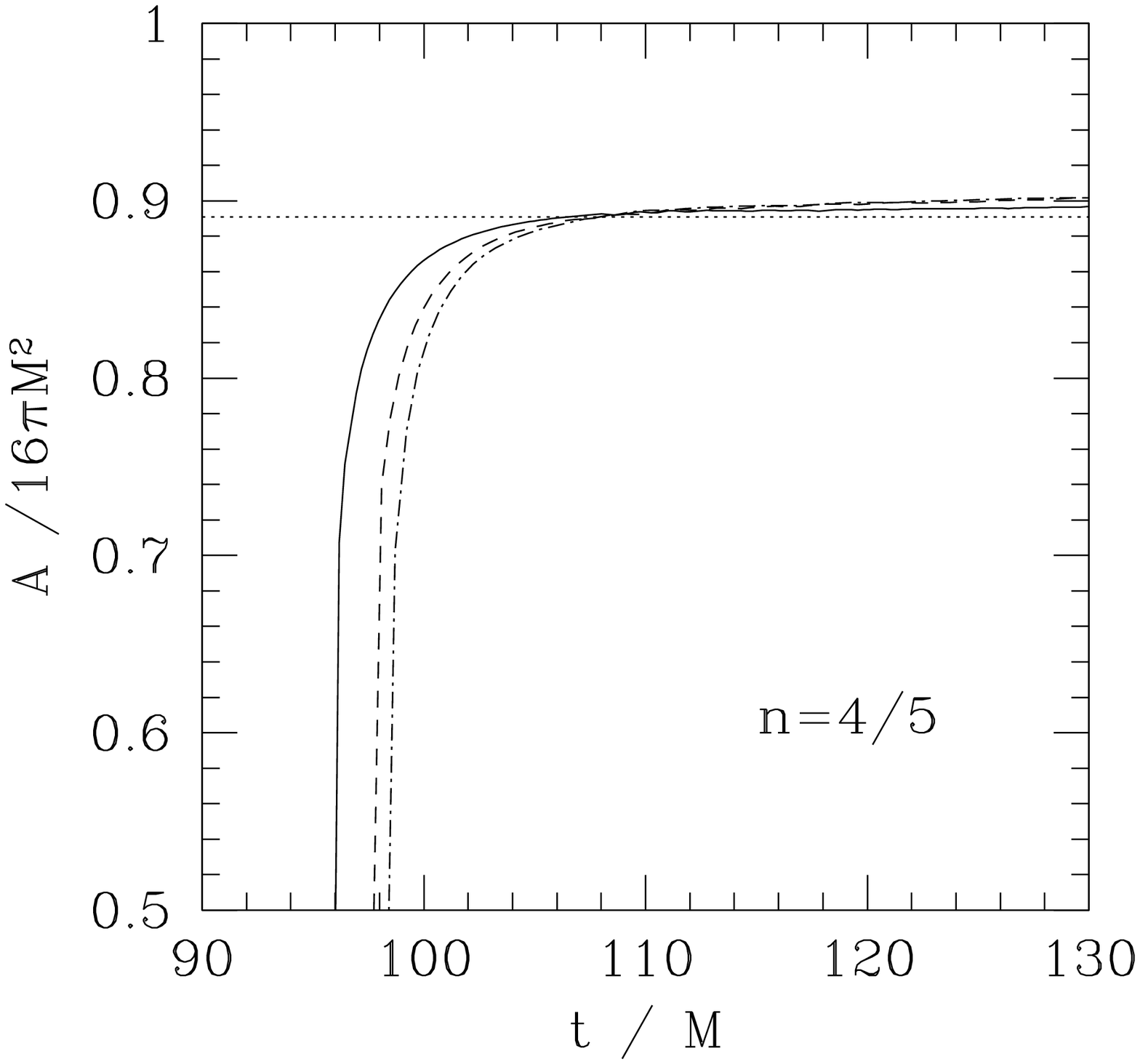}\\
\epsfxsize=2.6in
\leavevmode
(c)\epsffile{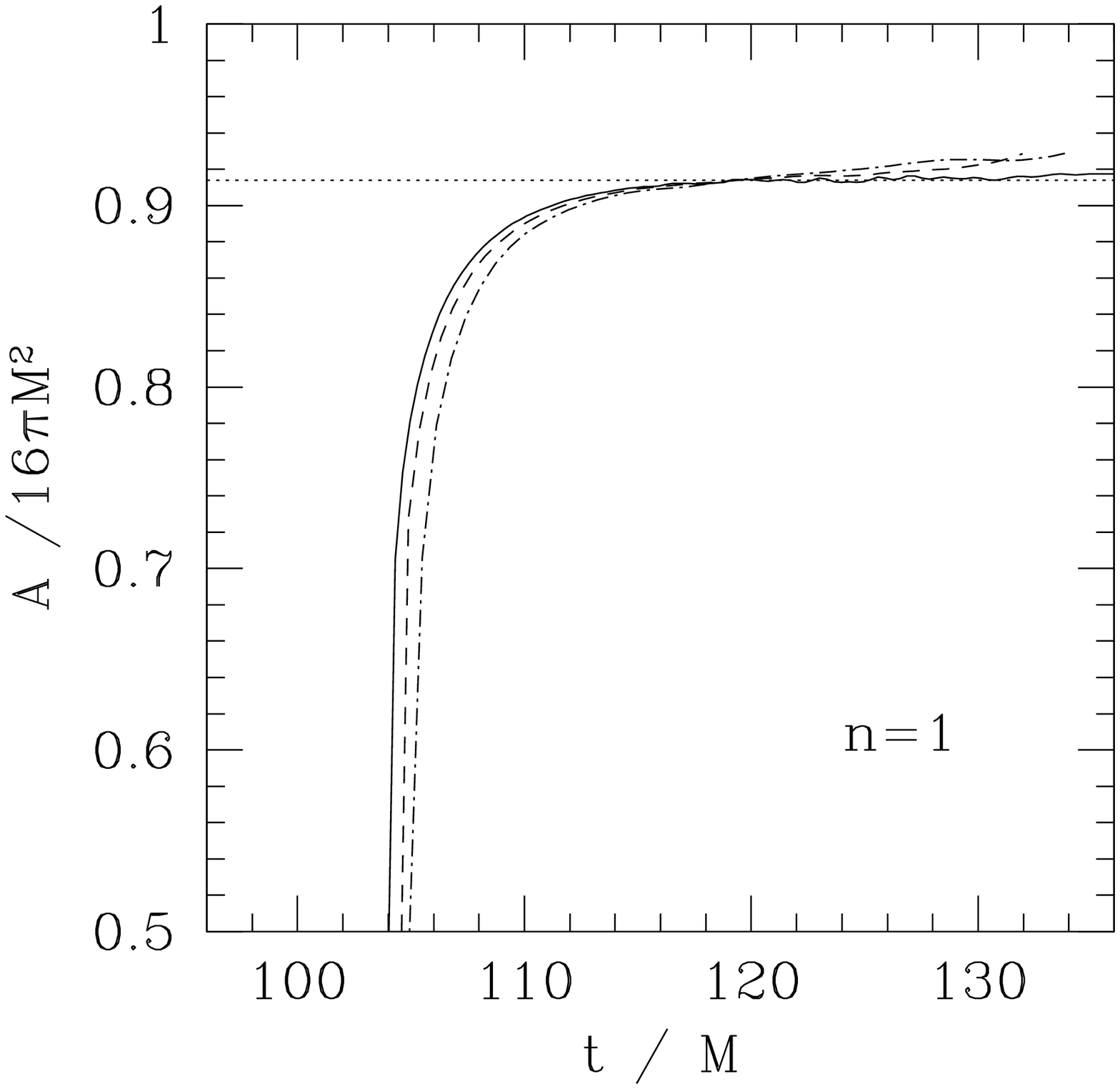}
\epsfxsize=2.6in
\leavevmode
~~(d)\epsffile{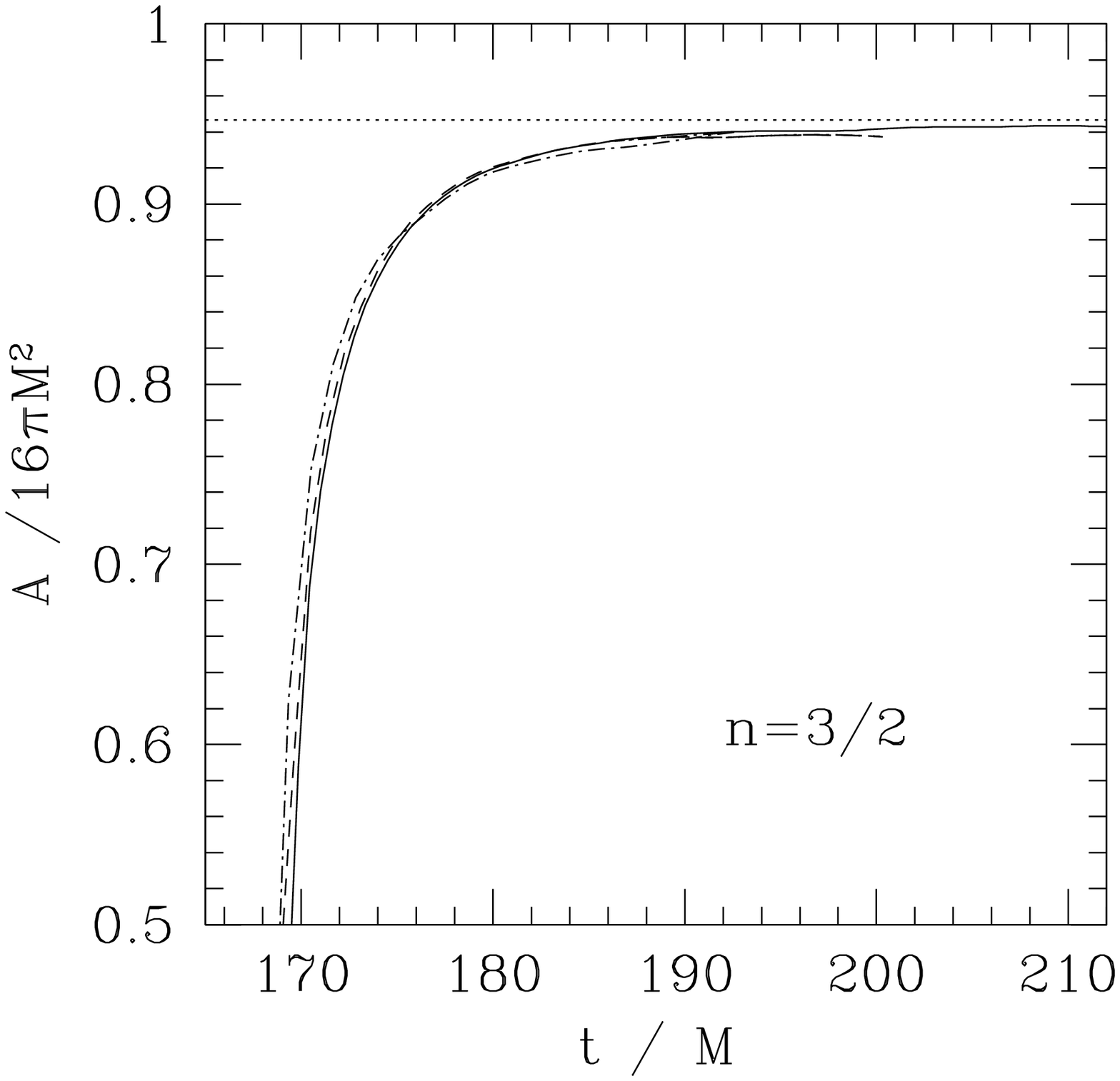}\\
\epsfxsize=2.6in
\leavevmode
(e)\epsffile{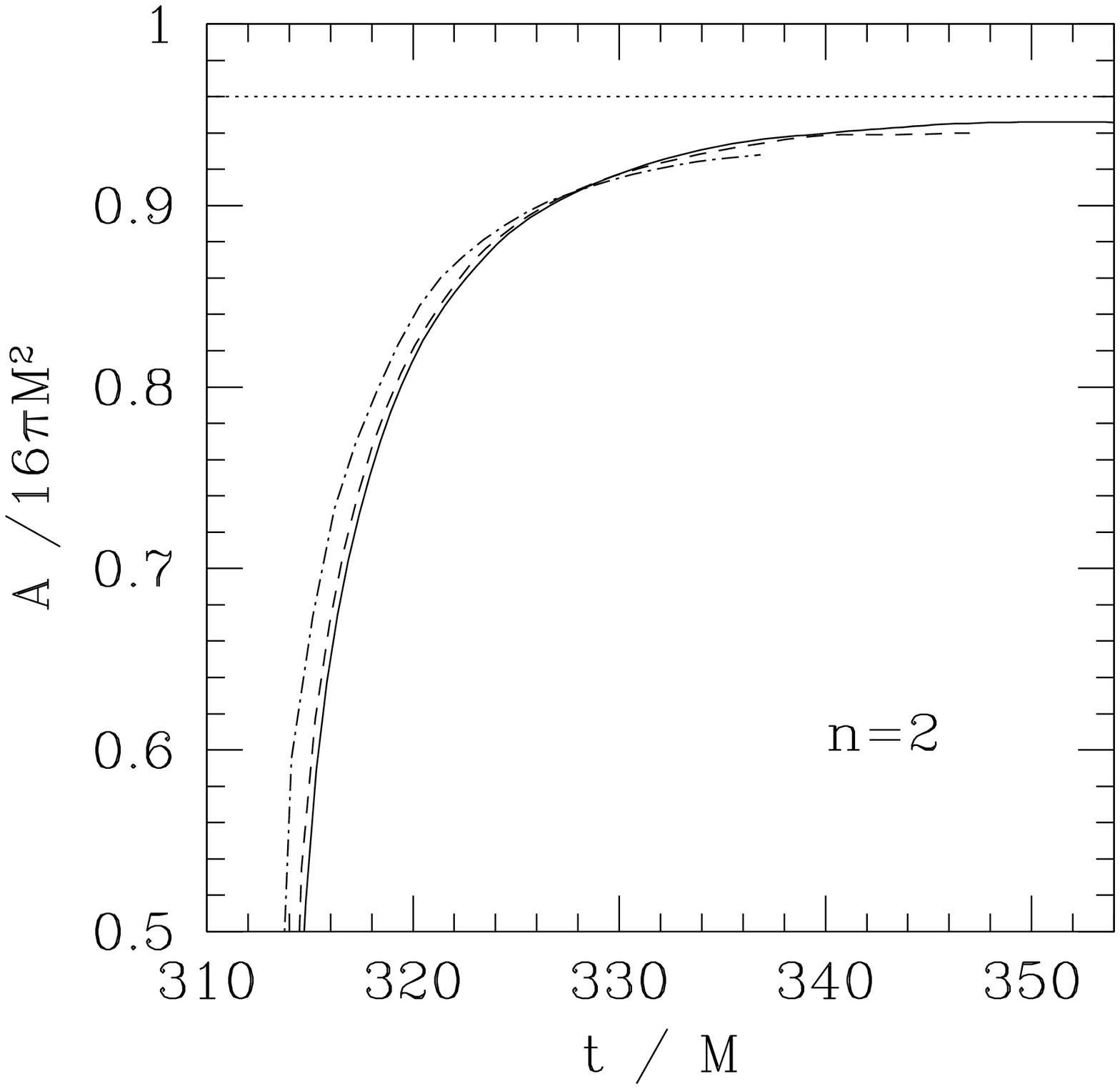}
\vspace*{-4mm}
\caption{Evolution of the area of the apparent horizon
in units of $16\pi M^2$ (i.e., the square of the
apparent horizon mass) 
for (a) $n=2/3$, (b) 4/5, (c) 1, (d) 3/2 and (d) 2. For 
(a)--(d), the solid, dashed and dotted-dashed curves are results
for $N=360$, 240, and 180, and for 
(e), the solid and dashed curves are results
for $N=480$, 360, and 240. 
The dotted horizontal lines denote the area of 
the event horizon for Kerr black holes of a given set of 
$J$ and $M$ for which we adopt the values of the initial conditions.
\label{FIG2}
}
\end{center}
\end{figure}

\begin{figure}[htb]
\vspace*{-4mm}
\begin{center}
\epsfxsize=3.5in
\leavevmode
\epsffile{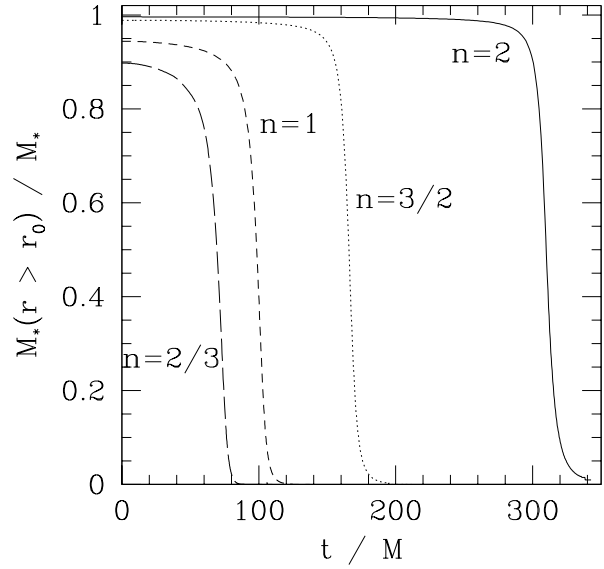}
\vspace*{-4mm}
\caption{Evolution of the total mass outside 
a coordinate radius $r_0$ for $n=2$ (solid curve),
3/2 (dotted curve), 1 (dashed curve),
and 2/3 (long-dashed curve).
$r_0$ is chosen as $\sim 0.6M$ which is approximately
equal to the asymptotic value of the 
coordinate polar-radius of the apparent horizon in the
present computations. 
To plot the curves, we choose the numerical results with $N=360$. 
The time is shown in unit of $M$. 
\label{FIG3}
}
\end{center}
\end{figure}

\begin{figure}[htb]
\vspace*{-4mm}
\begin{center}
\epsfxsize=3.5in
\leavevmode
\epsffile{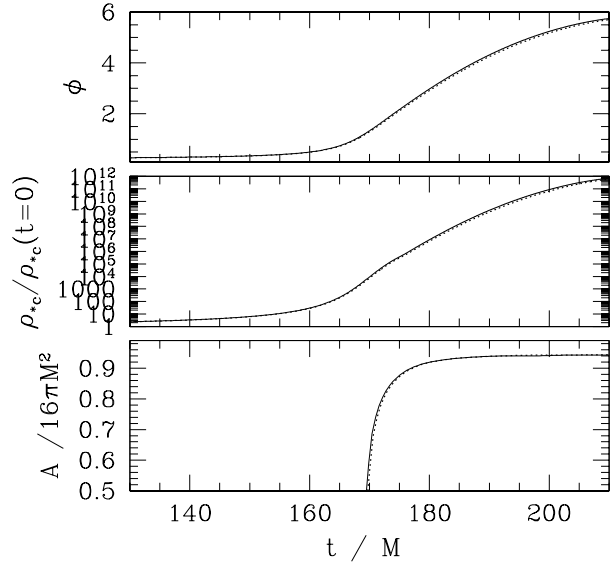}
\vspace*{-4mm}
\caption{Evolution of $\phi$ and $\rho_*$ at $r=0$ as well as 
$M_{\rm AH}^2$ for $n=3/2$ with $N=360$.
The solid and dotted curves denote the results by
the dynamical and AMD gauge conditions, respectively, and
they approximately coincide. 
\label{FIG4}
}
\end{center}
\end{figure}

\begin{figure}[htb]
\vspace*{-4mm}
\begin{center}
\epsfxsize=3.5in
\leavevmode
\epsffile{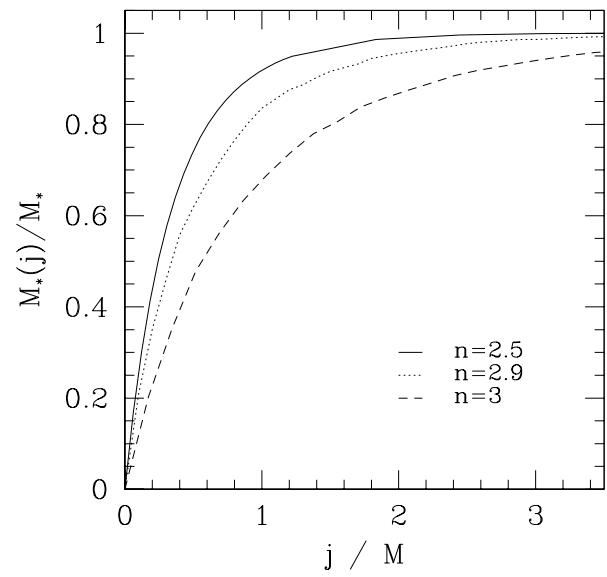}
\vspace*{-4mm}
\caption{The mass distribution as a function of the
specific angular momentum for $n=2.5$, 2.9, and 3. 
\label{FIG5}
}
\end{center}
\end{figure}


\end{document}